\documentclass[a4paper,12pt,twoside]{article}
\usepackage{latexsym}
\usepackage{ecltree}
\usepackage{epic}
\usepackage{eepic}
in25mm
\else\oddsidemargin25mm\evensidemargin25mm\marginparwidth25mm\fi
\begin{document}
\def\a{\alpha}
\def\b{\beta}
\def\g{\gamma}
\def\d{\delta}
\def\e{\epsilon}
\def\ve{\varepsilon}
\def\t{\theta}
\def\l{\lambda}
\def\m{\mu}
\def\n{\nu}
\def\pg{\pi}
\def\r{\rho}
\def\s{\sigma}
\def\t{\tau}
\def\c{\chi}
\def\p{\psi}
\def\o{\omega}
\def\G{\Gamma}
\def\D{\Delta}
\def\T{\Theta}
\def\L{\Lambda}
\def\Pg{\Pi}
\def\S{\Sigma}
\def\O{\Omega}
\def\pb{\bar{\psi}}
\def\cb{\bar{\chi}}
\def\lb{\bar{\lambda}}
\def\i{\imath}
\def\eq#1{(\ref{#1})}
\newcommand{\be}{\begin{equation}}
\newcommand{\ee}{\end{equation}}
\newcommand{\ba}{\begin{eqnarray}}
\newcommand{\ea}{\end{eqnarray}}
\newcommand{\ban}{\begin{eqnarray*}}
\newcommand{\ean}{\end{eqnarray*}}
\newcommand{\nn}{\nonumber}
\newcommand{\nin}{\noindent}

\begin{titlepage}
\vskip 0.5cm \hbox{July 2002} \vskip 2cm
\begin{center}
{\LARGE {On the construction of variant supergravities\\ in $D=11$, $D=10$}}\\
\vskip 1.5cm
  {\bf Silvia Vaul\`a} \\

\vskip 0.5cm {\small Dipartimento di Fisica, Politecnico di
Torino,\\
 Corso Duca degli Abruzzi 24, I-10129 Torino\\
and Istituto Nazionale di Fisica Nucleare (INFN) - Sezione di
Torino, Italy}\\
\vspace{6pt}
\end{center}
\vskip 3cm
\begin{abstract}
We construct with a geometric procedure the supersymmetry transformation laws and Lagrangian for all the ``variant'' $D=11$ and $D=10$ Type $IIA$ supergravities. We identify into our classification the  $D=11$ and $D=10$ Type $IIA$ ``variant'' theories first introduced by Hull performing $T$--duality transformation on both spacelike and timelike circles. We find in addition a set of $D=10$ Type $IIA$ ``variant'' supergravities that can not be obtained trivially from eleven dimensions compactifying on a circle.         
\end{abstract}

\end{titlepage}

\section{Introduction}
Variant supergravities were first introduced by Hull \cite{Hull:1998vg} considering $T$--duality transformations on timelike circles. Type $IIA$ and Type $IIB$ string theories are interchanged by $T$--duality on a spacelike circle while if the $T$--duality is performed on a timelike circle they transform respectively into the $IIB^*$ and $IIA^*$ theories, that are ``variant'' form of the conventional  Type $IIA$ and Type $IIB$ string theories, characterized by the presence of ghosts. From a supergravity point of view, this reflects into the fact that while the compactification on a spacelike circle of  Type $IIA$ and Type $IIB$ supergravities leads to the same nine dimensional theory,  the compactification on a timelike circle of  Type $IIA^*$ and Type $IIB^*$ supergravities leads to the same Euclidean nine dimensional theory.\\  Further developments \cite{Hull:1998ym} considered the possibility of perform more complicated $T$--duality on tori including both spacelike and timelike circles obtaining Type $II$ $D=10$ and $D=11$ variant supergravities with non Lorentzian space--time signature and often with ghost--like fields.\\   
In this paper we consider the problem of constructing such variant supergravities, particularly the variant $D=11$ and $D=10$ Type $IIA$ theories, from an algebraic point of view. In the present case, this can be done only for those signatures admitting reducible spinors, since we need no more than 32 supersymmetries in order to avoid fields with spin higher than two. In $D=11$ only the Majorana condition can be imposed and this is possible only for signatures $\r=\hat{s}-\hat{t}$=1, 7 (mod 8); for the non--chiral Type $IIA$, we can use only Majorana or Majorana--Weyl spinors, that in $D=10$ occur when the number of timelike dimensions is $t$=0, 2, 6 (mod 8) \cite{Kugo:1982bn}, \cite{Ferrara:2001sx}, \cite{Ferrara:2001dn}.\\ We know that conventional $D=11$ \cite{Cremmer:1978km}, \cite{Cremmer:1980ru}, \cite{Nicolai:1980kb}, \cite{D'Auria:nx} and $D=10$ Type $IIA$ \cite{Campbell:zc}, \cite{Bergshoeff:1981um} supergravities are local theories of the $OSp(1|32)$ supergroup, thus their fields form a real irreducible representation of $OSp(1|32)$, respectively on an eleven and ten dimensional Minkowski space--time. This implies, in the geometric (group--manifold or rheonomic) approach, that we are able to construct with such fields a Free Differential Algebra (FDA) \cite{cubo} (that is a set of generalized Maurer--Cartan equations closed under $d$--differentiation describing the supersymmetric vacuum configuration; this will be the starting point for the construction of the theory. From this point of view, if in $D=11$ or $D=10$ and in a given space--time signature we are able to write down one or more FDA involving a real representation of $OSp(1|32)$, in principle we can construct the corresponding supergravity. On the other hand, we know that $OSp(1|32)$ has only one real form. In reference \cite{Bergshoeff:2000qu} it is shown that in $D=11$ there is only one vector in the $OSp(1|32)$ algebra that can be identified with a real translational operator, then for each allowed signature there is only one real representation of $OSp(1|32)$; so we expect we are able to construct just one FDA for each signature and consequently only one supergravity. On the contrary in $D=10$ there are two possible identifications for the vector acting as a translational operator, then there are two real representation of $OSp(1|32)$ for each signature, and consequently we could in principle construct two distinct FDA and two supergravities.\\ In the present paper we will take as a starting point the construction of all possible FDA for each allowed signature. We will see how it is possible to identify the ghost content of each variant theory just from the FDA whit no need to construct the the theory. Then we outline a quite general procedure to construct a variant supergravity once the conventional one is known and apply it to the present case.\\  

The paper is organized as follows:

In Section 2 we construct all the possible FDA in eleven dimensions and derive the procedure which we will use to identify the structure of the $D=11$ variant supergravities. 

In Section 3 we repeat the same steps for the $D=10$ Type $IIA$ variant supergravities. 

In section 4 we perform explicitly the construction of the variant $D=11$ and $D=10$ Type $IIA$ supergravities from the conventional ones and give all the supersymmetry transformation laws and Lagrangians.  

In Section 5 we give our conclusions and in Appendix we discuss the consistence of the Majorana condition for different signatures and the definition of the Dirac conjugate spinors.

\section{Construction of $D=11$ Free Differential Algebras}
In this section consider the problem of constructing a $D=11$ supergravity with a non Lorentzian signature $(\hat{t},\hat{s})$, with \be\eta=\{\underbrace{+,+,+}_{\hat{t}\,times},\dots\underbrace{-,-,-}_{\hat{s}\,times}\}\label{eta}\ee that we will indicate with $M(\hat{t},\hat{s})$. First we observe that we can construct a $D=11$  supergravity only for those signatures allowing  Majorana spinors, since the number of real supercharges can not be higher than 32, in order to avoid the presence of fields whit spin higher than two. It is well known \cite{Kugo:1982bn} that this is possible only for the following signatures:
\begin{center}$M(1,10)\,\,\,\, M(2,9)\,\,\,\, M(5,6)\,\,\,\, M(6,5)\,\,\,\, M(9,2)\,\,\,\, M(10,1)$\end{center}   
The field content of the $D=11$ supermultiplet is given by the spin 2 vielbein field $V^{\hat{a}}_{\hat{\m}}$, the spin $\frac{3}{2}$ gravitino field $\p_{\hat{\m}}$ and by the spin 1 antisymmetric tensor field $C_{\hat{\m}\hat{\n}\hat{\r}}$; $\hat{a},\hat{b}=1,2\dots 11$ are flat indexes, $\hat{\m},\hat{\n}=1,2\dots 11$ are curved indexes.\\ The $M(\hat{t},\hat{s})$ supergravities are based on the following Free Differential Algebra (FDA)\cite{D'Auria:nx} describing the supersymmetric vacuum configuration. 
\ba
\label{dV}&&\mathcal{D}V^{\hat{a}}+\a\pb\G^{\hat{a}}\p=0\\
\label{do}&&\mathcal{R}^{\hat{a}\hat{b}}=0\\
\label{dr}&&\mathcal{D}\p=0\\
\label{dc}&&F=dC+\b\pb\G_{\hat{a}\hat{b}}\p V^{\hat{a}}V^{\hat{b}}=0
\ea
where $\mathcal{R}^{\hat{a}\hat{b}}=d\o^{\hat{a}\hat{b}}-\o^{\hat{a}}_{\,\,\hat{c}}\o^{\hat{c}\hat{b}}$ is the Lorentzian curvature and $\o^{\hat{a}\hat{b}}$ the spin connection, while $\mathcal{D}$ is the Lorentz derivative. $\G^{\hat{a}}$ are the eleven dimensional Gamma-matrices, satisfying the Clifford algebra\be\left\{\G^{\hat{a}},\G^{\hat{b}}\right\}=2\eta_{\hat{a}\hat{b}}\ee with $\eta_{\hat{a}\hat{b}}$ defined in equation \eq{eta}.\\
The closure under $d$-differentiation does not depend on $\a$ and $\b$ and relies on the following Fierz identity:
\be\label{fierz11}\pb\G^{\hat{a}}\p\pb\G_{\hat{a}\hat{b}}\p=0 \ee 
On the other hand, since the fields appearing in the $D=11$ multiplet are a real representation of $OSp(1|32)$ \cite{D'Auria:nx}, $\a$ and $\b$ must be chosen appropriately; in order to do that, we must analyze the reality of the currents appearing in the FDA, since it depends on the signature $(\hat{t},\hat{s})$ and on the way we define the Dirac conjugate spinor
\be\label{bar}\pb\equiv\p^{\dagger}G^{-1}\ee
We have two possible choices for the matrix $G$ in $D=t+s$
dimensions \cite{Regge:1983uu} in order to have $\pb\p$ transforming as a Lorentz scalar: the first choice corresponds to the product of all the timelike Gamma--matrices, that is:
\be \label{GI}G_I=\G^1\dots\G^t \ee
while the second choice corresponds to take the product of all
the spacelike Gamma--matrices:
\be\label{GII}G_{II}=\G^{D-s}\dots\G^{D}\ee 
Correspondingly the hermitian conjugate of the Gamma--matrices is given by:
\be\label{dagger1}\G^{a\dagger}=\eta\, G^{-1}\G^aG \ee
where the phase $\eta$ takes two different values according to the two choices of $G$, namely 
\be\label{dagger2}\eta_I=(-1)^{t-1};\,\,\,\,\,\eta_{II}=(-1)^s\ee
If we consider a current $j^{a_1\dots a_n}\equiv\lb\g^{a_1\dots a_n}\l$, where $\g^{a_1\dots a_n}$ is a totally antisymmetric product of
Gamma--matrices, we have \cite{Regge:1983uu}
\ba
&&(j^{a_1\dots a_n})^*=-\d\c j^{a_1\dots a_n}\nn\\
&&\c_I=(-1)^{\frac{1}{2}(t-n-1)(t-n)}\nn\\
\label{complex}&&\c_{II}=(-1)^{\frac{1}{2}(n-s-1)(n-s)}\ea 
where the subscripts $I$ and $II$ refer to the two possible choices for the matrix $G=\{G_I,\,\,G_{II}\}$ and $\d=\pm 1$ is the arbitrary
phase appearing in the convention one uses for the definition of the complex conjugate of the product of two spinors, namely \cite{Regge:1983uu}, \cite{VanProeyen:1999ni}
\be\label{coniugo}(\l\m)^*=\d\l^*\m^*=-\d\m^*\l^*\ee
Let us apply those considerations to the $D=11$ case, for the signatures we are interested in; we resume the result in Table \ref{11curr}. 
\vskip 0.5cm
\begin{table}[h]
\begin{center}
\begin{tabular}{|r|r|r|r|r|r|r|r|r|r|r|r|l|}
\hline
\,&\multicolumn{2}{r|}{$M(1,10)$}&\multicolumn{2}{r|}{$M(2,9)$}&\multicolumn{2}{r|}{$M(5,6)$}&\multicolumn{2}{r|}{$M(6,5)$}&\multicolumn{2}{r|}{$M(9,2)$}&\multicolumn{2}{r|}{$M(10,1)$}\\
\hline
\,&$\c_I$&$\c_{II}$&$\c_I$&$\c_{II}$&$\c_I$&$\c_{II}$&$\c_I$&$\c_{II}$&$\c_I$&$\c_{II}$&$\c_I$&$\c_{II}$\\
\hline
$\pb\G^a\p$&-1&+1&-1&-1&-1&+1&-1&-1&-1&+1&-1&-1\\
\hline
$\pb\G^{ab}\p$&+1&-1&-1&-1&+1&-1&-1&-1&+1&-1&-1&-1\\
\hline
\end{tabular}
\caption{Values of the phase $\c$ for $D=11$
currents}\label{11curr}
\end{center}
\end{table}
We can see that for each signature the choice I or II is equivalent, because of the arbitrary phase $\d$ (see Appendix). For the $M(1,10)$ theory we choose $\a=-\frac{i}{2}$ and we see that this leads to a real vielbein in equation \eq{dV} if we set, for instance, $\d=-1$ and $G=G_I$; consequently $\b$ should be a real number and we choose $\b=-\frac{1}{2}$, obtaining this way the following $M(1,10)$ FDA \cite{D'Auria:nx}:    
\ba
\label{dV+}&&\mathcal{D}V^{\hat{a}}-\frac{i}{2}\pb\G^{\hat{a}}\p=0\\
\label{do+}&&\mathcal{R}^{\hat{a}\hat{b}}=0\\
\label{dr+}&&\mathcal{D}\p=0\\
\label{dc+}&&F=dC-\frac{1}{2}\pb\G_{\hat{a}\hat{b}}\p V^{\hat{a}}V^{\hat{b}}=0
\ea
Looking at Table \ref{11curr} we see that, keeping $\d=-1$ and $G=G_I$, we can set $\a=-\frac{i}{2}$ for all signatures; on the other hand $\b$ must be real for signatures $M(1,10)$, $M(5,6)$, $M(9,2)$ an we set $\b=-\frac{1}{2}$, while it should be pure imaginary for signatures $M(10,1)$, $M(6,5)$, $M(2,9)$, in these cases we can choose $\b=-\frac{i}{2}$, obtaining this way a different FDA.
\ba
\label{dV++}&&\mathcal{D}V^{\hat{a}}-\frac{i}{2}\pb\G^{\hat{a}}\p=0\\
\label{do++}&&\mathcal{R}^{\hat{a}\hat{b}}=0\\
\label{dr++}&&\mathcal{D}\p=0\\
\label{dc++}&&\widetilde{F}=d\widetilde{C}-\frac{i}{2}\pb\G_{\hat{a}\hat{b}}\p V^{\hat{a}}V^{\hat{b}}=0
\ea
Obviously if we start from the FDA \eq{dV++}--\eq{dc++} and implement the rheonomic construction procedure we will obtain a theory which differs from the conventional one obtained from the  FDA \eq{dV+}--\eq{dc+}. By the way there is no need to construct from the very beginning the supersymmetry transformation laws and the Lagrangian for all the variant forms of $D=11$ supergravity; it is more useful to find some suitable redefinitions that transform the conventional $D=11$ supergravity into its variant forms.\\      
For signatures $M(10,1)$, $M(6,5)$, $M(2,9)$ if we keep $\b=-\frac{1}{2}$, that is we start from the same FDA \eq{dV+}--\eq{dc+} of the conventional theory, we would work with a pure imaginary super covariant field strength $F$. We know that this do not correspond to a real representation but the closure of the FDA depends neither on the space-time signature, nor on the reality of the fields; the Fierz identity \eq{fierz11} follows from the absence of some threelinear fermionic representations of $SO(11)$, and of all its non compact forms $SO(\hat{t},\hat{s})$. The whole rheonomic procedure of constructing the supersymmetry transformation laws from the FDA (see for instance \cite{cubo}) does not depend on the space-time signature and the reality of the fields; once the supersymmetry transformation laws are given, is easy to see that also the Einstein term and the bosonic kinetic terms of a supersymmetric Lagrangian are certainly  not depending on these latter. This means that if we keep $\b=-\frac{1}{2}$ for all $M(\hat{t},\hat{s})$ theories, we have formally the same FDA for them all, consequently we will get formally the same supersymmetric transformation laws and the same kinetic Lagrangian. At this point one has just to recall that for $M(10,1)$, $M(6,5)$, $M(2,9)$ theories the field $F$ is pure imaginary and it must be expressed in terms of a real field $\widetilde{F}$, that is $F=-i\widetilde{F}$. Consequently, since the $M(1,10)$ theory is ghost free, the kinetic term of the field $F$ undergoes a change of sign; hence for $M(10,1)$, $M(6,5)$, $M(2,9)$ theories the real field $\widetilde{F}$ is a ghost.\\
This is the same result conjectured by Hull \cite{Hull:1998ym},\cite{Hull:1999mr} . Since we considered the problem with a constructive geometric approach, we can conclude that these are the only $M(\hat{t},\hat{s}
)$ theories one can construct and therefore we are able to give a simple recipe to determine their supersymmetry transformation laws and Lagrangian as will be discussed in Section 4. This algebraic argument is related to the one used in reference \cite{Bergshoeff:2000qu}: there exists only one real section of the complex $OSp(1|32)$ superalgebra and in $D=11$ there is only one way to identify the generators acting as space-time translations, so the $D=11$ realization of the real section of the algebra is unique for each signature. On the other hand, in $D=10$ there are two possibilities of identifying the generators of space-time translations, and this doubles the theories for each signature. This will be discussed in the next section.     

\section{Construction of $D=10$ Type $IIA$ Free Differential Algebras}

We want now to make the same analysis for $D=10$ Type $IIA$ supergravity for various signatures, that we will indicate as $IIA(t,s)$ theories. As in the previous $D=11$ case, we can consider just the signatures admitting Majorana spinors, that is:     
\begin{center}$IIA(0,10)\,\,\,\, IIA(1,9)\,\,\,\, IIA(2,8)\,\,\,\, IIA(4,6)\,\,\,\, IIA(5,5)$\end{center}                   
and their mirrors $t\leftrightarrow s$ (for signatures $IIA(1,9)$, $IIA(9,1)$ and $IIA(5,5)$ we have Majorana-Weyl spinors, but now we prefer to use Majorana spinors $\p=\p_L+\p_R$ for homogeneity of notation).\\ 
It is well known that $Type-IIA(1,9)$ supergravity can be obtained from $M(1,10)$ supergravity compactifying on a space-like circle, and recently was found \cite{Hull:1998ym} that the $M(\hat{t},\hat{s})$ and the $IIA(t,s)$ supergravities are linked each other via $T-duality$ on tori including timelike circles. In particular it was found \cite{Hull:1998ym} that reducing $M(\hat{t},\hat{s})$ theories on spacelike or timelike circles gives rise to $IIA(t,s)$ theories according to the following scheme:
\begin{center}
\begin{bundle}{$M(1,10)$}\chunk{\begin{bundle}{time}\chunk{$IIA(0,10)$}\end{bundle}}\chunk{\begin{bundle}{space}\chunk{$IIA(1,9)$}\end{bundle}}\end{bundle}\hspace{1cm}\begin{bundle}{$M(2,9)$}\chunk{\begin{bundle}{time}\chunk{$IIA^*(1,9)$}\end{bundle}}\chunk{\begin{bundle}{space}\chunk{$IIA(2,8)$}\end{bundle}}\end{bundle}\hspace{1cm}\begin{bundle}{$M(5,6)$}\chunk{\begin{bundle}{time}\chunk{$IIA(4,6)$}\end{bundle}}\chunk{\begin{bundle}{space}\chunk{$IIA(5,5)$}\end{bundle}}\end{bundle}
\vskip 1cm
\begin{bundle}{$M(10,1)$}\chunk{\begin{bundle}{time}\chunk{$IIA(9,1)$}\end{bundle}}\chunk{\begin{bundle}{space}\chunk{$IIA(10,0)$}\end{bundle}}\end{bundle}\hspace{1cm}\begin{bundle}{$M(9,2)$}\chunk{\begin{bundle}{time}\chunk{$IIA(8,2)$}\end{bundle}}\chunk{\begin{bundle}{space}\chunk{$IIA^*(9,1)$}\end{bundle}}\end{bundle}\hspace{1cm}\begin{bundle}{$M(6,5)$}\chunk{\begin{bundle}{time}\chunk{$IIA^*(5,5)$}\end{bundle}}\chunk{\begin{bundle}{space}\chunk{$IIA(6,4)$}\end{bundle}}\end{bundle}
\end{center}
\be\label{diag}\ee		
The $IIA^*$ theories are distinct from the ones with the same space--time signature \cite{Hull:1998ym}, \cite{Bergshoeff:2000qu}. 
When the reduction is performed on the $D=11$ FDA, the gamma--matrices reduce according to the following rules:
\ba
&&\G^{\hat{a}}\longrightarrow\g^a\,\,\,\,a,\hat{a}=1\dots10\nn\\
\label{gred}&&\G^{11}\longrightarrow\g^{11}\equiv\phi\g^1\g^2\dots\g^{10}
\ea 
The phase $\phi$ must be fixed according to the square in eleven dimensions: $(\G^{11})^2=1$ if we compactify on a timelike circle, $(\G^{11})^2=-1$ on a spacelike circle. According to our choice for the signature we have:
\be\label{square}(\g^{11})^2=\phi^2(-1)^{s+1}\ee 
The relevant thing is how the definition of the Dirac conjugate spinor $\pb=\p^{\dagger}G^{-1}$ changes reducing from $D=11$ to $D=10$; let us consider the various cases.\\
If the Dirac conjugate spinor in $D=11$ is defined using $G=G_I=\G^1\dots\G^{\hat{t}}$ there are two possible cases: if we reduce on a spacelike circle $\G^{11}$ does not appear in the definition of $\pb$ and hence the Dirac conjugate spinor reduces to $D=10$ as\be\pb=\p^{\dagger}(G_I^{(11)})^{-1};\quad G_I^{(11)}=\G^1\dots\G^{\hat{t}}\quad\stackrel{D=10}\longrightarrow\quad\pb=\p^{\dagger}(G_I^{(10)})^{-1};\quad G_I^{(10)}=\g^1\dots\g^t\ee On the other hand, if we reduce on a timelike circle  $\G^{11}$ appears in the definition of $\pb$, since $G=G_I=\G^1\dots\G^{\hat{t}}=\G^1\dots\G^{\hat{t}-1}\G^{11}$; recalling definition \eq{gred} for the eleventh Gamma--matrix, we have that\be\pb=\p^{\dagger}(G_I^{(11)})^{-1};\quad G_I^{(11)}=\G^1\dots\G^{\hat{t}-1}\G^{11}\quad\stackrel{D=10}\longrightarrow\quad\pb=\xi\p^{\dagger}(G_{II}^{(10)})^{-1};\quad G_{II}^{(10)}=\g^{10-s}\dots\g^{10}\ee
where $\xi$ is a phase depending on $\phi$ and on the number of spacelike dimensions that has to be determined for each case.\\      
On the other hand, if in $D=11$ the Dirac conjugate spinor is defined using $G=G_{II}$, the reduction on a timelike circle leaves the definition unchanged, while a reduction on a spacelike circle leads to a definition in $D=10$ with $G=G_I$, up to a phase.\\ This is a general feature of reduction from $D=2n+1\longrightarrow D-1=2n$: if we reduce on a spacelike circle we obtain a theory in $D-1$ dimensions formulated with $\pb=\p^{\dagger}G_I^{-1}$, if we reduce on a timelike circle we obtain a theory formulated with $\pb=\p^{\dagger}G_{II}^{-1}$, and this does not depend on the definition of $\pb$ in $D$ dimensions\footnote{This also happens while constructing (if possible) $Anti\,de\,Sitter$ and $de\,Sitter$ supergravities in even dimensions \cite{D'Auria:2002fh}}.\\
We will not perform here the explicit reduction to ten dimensions for the various signatures; it is more useful to start from the very beginning and construct all the possible $IIA(t,s)$ with the geometric approach considering the appropriate $D=10$ FDA.\\ However the previous discussion was important to realize that we must take into account both type of theories, with Dirac conjugate spinors defined with $G=G_I$ and $G=G_{II}$ and will help us to identify the theories we are going to construct with the ones classified by Hull in reference \cite{Hull:1998ym}, summarized in the diagram \eq{diag}.
\newpage
We take as starting point the FDA of the already known $IIA(1,9)$ theory, obtained compactifying the FDA of $M(1,10)$ supergravity \cite{D'Auria:nx} on a spacelike circle:  
\ba
&&\label{dv-}\mathcal{D}V^a-\frac{i}{2}\pb\g^a\p=0\\
&&\label{do-}\mathcal{R}^{ab}=0\\
&&\label{dr-}\mathcal{D}\p=0\\
&&\label{da-}G=dA-\frac{i}{2}\pb\g^{11}\p=0\\
&&\label{db-}H=dB-\frac{1}{2}\pb\g^a\g^{11}\p V_a=0\\
&&\label{dc-}F=dC-\frac{1}{2}\pb\g^{ab}\p V_aV_b+iB\pb\g^{11}\p=0
\ea   
Besides the graviton, described by the vielbein $V^a_{\m}$, the spin $\frac{3}{2}$ gravitino $\p_{\m}$, the one-form field $A_{\m}$, the two-form field $B_{\m\n}$ and the three-form $C_{\m\n\r}$, the $D=10$, $OSp(1|32)$ multiplet contains the spin $\frac{1}{2}$ dilatino $\c$ and the scalar dilaton $\s$, that are zero in the vacuum.\\
In order to determine the structure of all the $IIA(t,s)$ theories, we must analyze the reality properties of the currents involved in the FDA; it is useful to define 
\be\label{prodotto}\g=\g^1\g^2\dots\g^{10}\ee
so that the matrix $\g^{11}$ appearing in the $IIA(1,9)$ FDA \eq{dv-}--\eq{dc-} is defined as 
\be\label{g11}\g^{11}=i\g\,\,\,\,\,\,(\g^{11})^2=-1\ee
The reality properties of the $D=10$ currents for the various signatures $(t,s)$, are resumed in Table \ref{10curr}:    
\vskip 0.5cm
\begin{table}[h]\begin{center}\begin{tabular}{|r|r|r|r|r|r|r|r|r|r|l|}
\hline
$(t,s)$&\multicolumn{2}{r|}{$(0,10)$}&\multicolumn{2}{r|}{$(1,9)$}&\multicolumn{2}{r|}{$(2,8)$}&\multicolumn{2}{r|}{$(4,6)$}&\multicolumn{2}{r|}{$(5,5)$}\\
\hline
\,&$\c_I$&$\c_{II}$&$\c_I$&$\c_{II}$&$\c_I$&$\c_{II}$&$\c_I$&$\c_{II}$&$\c_I$&$\c_{II}$\\
\hline
$\pb\g^a\p$&+1&+1&-1&-1&-1&-1&+1&+1&-1&-1\\
\hline
$\pb\g^{ab}\p$&+1&-1&+1&-1&-1&+1&+1&-1&+1&-1\\
\hline
$\pb\g^{a}\g\p$&+1&+1&-1&-1&-1&-1&+1&+1&-1&-1\\
\hline
$\pb\g\p$&+1&-1&+1&-1&-1&+1&+1&-1&+1&-1\\
\hline
$(t,s)$&\multicolumn{2}{r|}{$(10,0)$}&\multicolumn{2}{r|}{$(9,1)$}&\multicolumn{2}{r|}{$(8,2)$}&\multicolumn{2}{r|}{$(6,4)$}&\multicolumn{2}{r|}{\,}\\
\hline
\,&$\c_I$&$\c_{II}$&$\c_I$&$\c_{II}$&$\c_I$&$\c_{II}$&$\c_I$&$\c_{II}$&\multicolumn{2}{r|}{\,}\\
\hline
$\pb\g^a\p$&-1&-1&-1&-1&+1&+1&-1&-1&\multicolumn{2}{r|}{\,}\\
\hline
$\pb\g^{ab}\p$&-1&+1&+1&-1&+1&-1&-1&+1&\multicolumn{2}{r|}{\,}\\
\hline
$\pb\g^{a}\g\p$&-1&-1&-1&-1&+1&+1&-1&-1&\multicolumn{2}{r|}{\,}\\
\hline
$\pb\g\p$&-1&+1&+1&-1&+1&-1&-1&+1&\multicolumn{2}{r|}{\,}\\
\hline
\end{tabular}
\caption{Values of the phase $\c$ for $D=10$ currents}\label{10curr}\end{center}\end{table}
We can easily see, recalling definition \eq{g11}, that the FDA \eq{dv-}--\eq{dc-} is then defined with real fields if we choose $\d=-1$ and $G=G_I$. As we did in $D=11$, we want now to find a relation between the already known supersymmetry transformation laws and Lagrangian of $IIA(1,9)$ supergravity and the other we are going to construct. The reasoning is the same: since the procedure of constructing the theory is dependent neither on the signature, nor on the definition of the Dirac conjugate, nor on the reality of the fields, we will keep the FDA \eq{dv-}--\eq{dc-} for all the theories and consequently we will get formally the same solution. Next, we will redefine the fields that are pure imaginary in terms of real ones: this way, since the $IIA(1,9)$ theory is ghost free, whenever an imaginary field occurs, it will be associated to a real ghost--like field. With this analysis we are able to determine which fields are ghost in each theory just from an algebraic constructive approach. When scalar fields are involved, some more considerations are needed to understand if the scalar will be a ghost or if the redefinition $\c\rightarrow i\c$ on the spin $\frac{1}{2}$ field is needed; it is easy to see that the dilaton can never become a ghost, as explained in the Appendix. We must also recall that, when doing explicit calculation starting from the $IIA(1,9)$ FDA \eq{dv-}--\eq{dc-}, we make use of the fact that $(\g^{11})^2=-1$, where $\g^{11}=i\g$; hence, if we want to use the same FDA for all the theories, we must define case by case a matrix $\g^{11}$ which squares to -1. Recalling equation \eq{square}, we see that $(\g)^2=-1$ if $s$ is even, hence, in this case, we should define $\g^{11}=\g$. Consequently one has to take into account the difference of a factor ''$-i$'' while evaluating the reality of the currents for theories with an even number of spacelike dimensions. Last, since it is convenient to leave unchanged the torsion equation \eq{dv-}, it is useful to set $\d=1$ for signatures $(0,10)$, $(4,6)$, $(8,2)$ and $\d=-1$ for the others as we can see from Table \ref{10curr}. For sake of clearness, let us first identify the theories occurring in reference \cite{Hull:1998ym} with the ones appearing in Table \ref{10curr}. Remembering that for a theory obtained compactifying on a spacelike circle $G=G_I$ while for a theory obtained compactifying on a timelike circle $G=G_{II}$ we have the following set of identifications:\\
\ba
&&(t,s)=(1,9):\quad G=G_I\longrightarrow IIA(1,9);\quad G=G_{II}\longrightarrow IIA^*(1,9)\nn\\
\nn\\
&&(t,s)=(9,1):\quad G=G_I\longrightarrow IIA^*(9,1);\quad G=G_{II}\longrightarrow IIA(9,1)\nn\\
\nn\\
&&(t,s)=(5,5):\quad G=G_I\longrightarrow IIA(5,5);\quad G=G_{II}\longrightarrow IIA^*(5,5)\label{doppio}\\
\nn\\
&&(t,s)=(0,10):\quad G=G_{II}\longrightarrow IIA(0,10)\nn\\
\nn\\
&&(t,s)=(10,0):\quad G=G_I\longrightarrow IIA(10,0)\nn\\
\nn\\
&&(t,s)=(2,8):\quad G=G_I\longrightarrow IIA(2,8)\nn\\  
\nn\\
&&(t,s)=(8,2):\quad G=G_{II}\longrightarrow IIA(8,2)\nn\\
\nn\\
&&(t,s)=(4,6):\quad G=G_{II}\longrightarrow IIA(4,6)\nn\\
\nn\\
&&(t,s)=(6,4):\quad G=G_{I}\longrightarrow IIA(6,4)\label{singolo}
\ea
We can observe that for the first group of signatures \eq{doppio} we have both the $IIA$ and the $IIA^*$ theories, while for the second group \eq{singolo} we have just one theory for each signature; all together, these are the only theories that can be obtained from $D=11$ compactifying on a spacelike or on a timelike circle, according to reference \cite{Hull:1998ym}.\\  
On the other hand, as we can see from Table \ref{10curr} these are not the only $IIA$ theories one can construct; for each signature $(t,s)$ we have always two distinct FDA formulated respectively with $G=G_I$ and $G=G_{II}$ corresponding to two different choices of the generators of the translations as explained in reference \cite{Bergshoeff:2000qu}; it is clear from Table \ref{10curr} that they are distinct theories with a different content of ghost. It is also evident that they can not be obtained from $D=11$ supergravity compactifying on just one circle.\\ We will indicate the theories missing from Hull's classification as $IIA'$ according to Van Proeyen's and Bergshoeff's notation \cite{Bergshoeff:2000qu} and they can be identified as follows\footnote{The authors of \cite{Bergshoeff:2000qu} propose a redefinition that identifies the algebra underlying the theories \eq{manca} with the one of \eq{singolo}; with this formulation it is not evident how to identify the \eq{manca} with the \eq{singolo} supergravities.}:\\
\ba
&&(t,s)=(0,10):\quad G=G_{I}\longrightarrow IIA'(0,10)\nn\\
\nn\\
&&(t,s)=(10,0):\quad G=G_{II}\longrightarrow IIA'(10,0)\nn\\
\nn\\
&&(t,s)=(2,8):\quad G=G_{II}\longrightarrow IIA'(2,8)\nn\\  
\nn\\
&&(t,s)=(8,2):\quad G=G_{I}\longrightarrow IIA'(8,2)\nn\\
\nn\\
&&(t,s)=(4,6):\quad G=G_{I}\longrightarrow IIA'(4,6)\nn\\
\nn\\
&&(t,s)=(6,4):\quad G=G_{II}\longrightarrow IIA'(6,4)\label{manca}
\ea
This scenario is similar to what happens in $D=6$ for the complex $F(4)^t$ superalgebra \cite{D'Auria:2002fh}: for $t=1,2$ \cite{VanProeyen:1999ni}, \cite{Ferrara:2001sx} we have two different real sections, that correspond to a $de\,Sitter$ and an $Anti\,de\,Sitter$ vacuum respectively. If we consider their contraction $m=0$ to a Poincar\'e vacuum, they are still two different theories; the contraction coming from $AdS$ is ghost free, while in the contraction from $dS$ the spin 1 fields $F^{\L}$ are ghosts, as in the uncontracted theory. In any case, the two FDA are characterized by a different definition of the Dirac conjugate spinor, with $G=G_I$ and $G=G_{II}$ respectively.\\
At this point we can simplify the content of Table \ref{10curr} taking into account all the previous considerations. In Table \ref{10curr2} we have grouped the theories with the same characteristics and we indicate the reality of each current. For each group of theories the ghost-like fields are the ones defined with currents whose reality differs from the one of the $IIA(1,9)$ theory, that is known to be ghost free. In Table \ref{ghostfields} we explicitly write the sign of the kinetic terms for each group of theories; we write a ''+'' when the sign is the same as the $IIA(1,9)$ theory and a ''--'' if it differs.\\ Note that, due to the structure of the supercovariant field strength \eq{dc-} the only possible combination of ghost and no--ghost fields are those of Table \ref{ghostfields}.
\vskip 0.5cm
\begin{table}[h]
\begin{center}
\begin{tabular}{|r|r|r|r|r|l|}
\hline
\,&$IIA(1,9)$&$IIA^*(1,9)$&$IIA(0,10)\,IIA'(10,0)$&$IIA(10,0)\,IIA'(0,10)$\\
\,&$IIA(5,5)$&$IIA^*(5,5)$&$IIA(4,6)\,IIA'(6,4)$&$IIA(6,4)\,IIA'(4,6)$\\
\,&$IIA^*(9,1)$&$IIA(9,1)$&$IIA(8,2)\,IIA'(2,8)$&$IIA(2,8)\,IIA'(8,2)$\\
\hline
$\pb\g^a\p$&imaginary&imaginary&imaginary&imaginary\\
\hline
$\pb\g^{ab}\p$&real&imaginary&real&imaginary\\
\hline
$\pb\g^{a}\g^{11}\p$&real&real&imaginary&imaginary\\
\hline
$\pb\g^{11}\p$&imaginary&real&real&imaginary\\
\hline
\end{tabular}
\caption{Reality of the currents for $D=10$ Type $IIA$ supergravities}
\label{10curr2}
\end{center}
\end{table}
\begin{table}[h]
\begin{center}
\begin{tabular}{|r|r|r|r|r|l|}
\hline
\,&$IIA(1,9)$&$IIA^*(1,9)$&$IIA(0,10)\,IIA'(10,0)$&$IIA(10,0)\,IIA'(0,10)$\\
\,&$IIA(5,5)$&$IIA^*(5,5)$&$IIA(4,6)\,IIA'(6,4)$&$IIA(6,4)\,IIA'(4,6)$\\
\,&$IIA^*(9,1)$&$IIA(9,1)$&$IIA(8,2)\,IIA'(2,8)$&$IIA(2,8)\,IIA'(8,2)$\\
\hline
$F$&+&--&+&--\\
\hline
$H$&+&+&--&--\\
\hline
$G$&+&--&--&+\\
\hline
\end{tabular}
\caption{Presence of ghosts in $D=10$ Type $IIA$ supergravities}
\label{ghostfields}
\end{center}
\end{table}

\section{The supersymmetric transformation laws\\ and Lagrangians}
\subsection{The $D=11$ supergravities}
In this section we will write explicitly the supersymmetry transformation rules and the relevant terms of the $D=11$ and  $D=10$ Type $IIA$ variant  supergravities Lagrangian, applying the redefinitions discussed in the previous sections.\\
As explained in Section 2, the $M(1,10)$, $M(5,6)$, $M(9,2)$ theories have the same structure, that is the well known conventional $D=11$ supergravity \cite{Cremmer:1978km} to which we refer from now on\footnote {There are some small differences in the conventions and normalization with respect to reference \cite{D'Auria:nx} to which we referred in Section 2, but this is irrelevant for our purposes}. 
The supersymmetry transformation laws are:
\ba
\label{uno11ng}&&\d V^{\hat{a}}_{\hat{\m}}=-i\overline{\ve}\G^{\hat{a}}\p_{\hat{\mu}}\\
\label{due11ng}&&\d\p_{\hat{\m}}=\mathcal{D}_{\hat{\m}}\ve+\frac{i}{144}(\G_{\hat{\m}}^{\phantom{a}\hat{\n}\hat{\r}\hat{\s}\hat{\l}}-8\d^{\hat{\n}}_{\hat{\m}}\G^{\hat{\r}\hat{\s}\hat{\l}})F_{\hat{\n}\hat{\r}\hat{\s}\hat{\l}}\ve\\
\label{tre11ng}&&\d C_{\hat{\m}\hat{\n}\hat{\r}}=\frac{3}{2}\overline{\ve}\G_{\hat{\m}\hat{\n}}\p_{\hat{\r}}
\ea
Furthermore the Lagrangian is given by:
\ba
(det\,V)^{-1}\mathcal{L}&=&-\frac{1}{4}\mathcal{R}-\frac{i}{2}\pb_{\hat{\m}}\G^{\hat{\m}\hat{\n}\hat{\r}}\mathcal{D}_{\hat{\n}}\p_{\hat{\r}}-\frac{1}{48}\mathcal{F}_{\hat{\m}\hat{\n}\hat{\r}\hat{\s}}\mathcal{F}^{\hat{\m}\hat{\n}\hat{\r}\hat{\s}}+\frac{1}{192}\left(\pb_{\hat{\m}}\G^{\hat{\m}\hat{\n}\hat{\r}\hat{\s}\hat{\l}\hat{\pi}}\p_{\hat{\n}}+12\pb^{\hat{\r}}\G^{\hat{\s}\hat{\l}}\p^{\hat{\pi}}\right)\nn\\
&&\left(\mathcal{F}_{\hat{\r}\hat{\s}\hat{\l}\hat{\pi}}+F_{\hat{\m}\hat{\n}\hat{\r}\hat{\s}}\right)+\frac{2}{(12)^{4}}(det\,V)^{-1}\e^{\hat{\a}\hat{\b}\hat{\g}\hat{\d}\hat{\l}\hat{\m}\hat{\n}\hat{\pi}\hat{\s}\hat{\t}\hat{\o}}\mathcal{F}_{\hat{\a}\hat{\b}\hat{\g}\hat{\d}}\mathcal{F}_{\hat{\l}\hat{\m}\hat{\n}\hat{\pi}}C_{\hat{\s}\hat{\t}\hat{\o}}\label{L11ng}
\ea
where we have defined according to reference \cite{Cremmer:1978km} the field strength $\mathcal{F}_{\hat{\m}\hat{\n}\hat{\r}\hat{\s}}=4\partial_{[\hat{\m}}C_{\hat{\n}\hat{\r}\hat{\s}]}$ and the super covariant field strength $F_{\hat{\m}\hat{\n}\hat{\r}\hat{\s}}=\mathcal{F}_{\hat{\m}\hat{\n}\hat{\r}\hat{\s}}-3\pb_{\hat{\m}}\G_{\hat{\n}\hat{\r}}\p_{\hat{\s}}$.\\For the $M(10,1)$, $M(6,5)$, $M(2,9)$ theories, the supersymmetry transformation rules and the Lagrangian are formally the same as \eq{uno11ng}--\eq{L11ng} but the super covariant field strength $F$ and the field strength $\mathcal{F}$ are pure imaginary. Expressing them in terms of real objects \be F=-i\widetilde{F};\quad\mathcal{F}=-i\widetilde{\mathcal{F}};\quad C=-i\widetilde{C}\ee  we thus obtain the supersymmetry transformation laws and the Lagrangian for the $M(10,1)$, $M(6,5)$, $M(2,9)$ theories.\\ The supersymmetry transformation laws are:  
\ba
\label{uno11g}&&\d V^{\hat{a}}_{\hat{\m}}=-i\overline{\ve}\G^{\hat{a}}\p_{\hat{\mu}}\\
\label{due11g}&&\d\p_{\hat{\m}}=\mathcal{D}_{\hat{\m}}\ve+\frac{1}{144}(\G_{\hat{\m}}^{\phantom{a}\hat{\n}\hat{\r}\hat{\s}\hat{\l}}-8\d^{\hat{\n}}_{\hat{\m}}\G^{\hat{\r}\hat{\s}\hat{\l}})\widetilde{F}_{\hat{\n}\hat{\r}\hat{\s}\hat{\l}}\ve\\
\label{tre11g}&&\d\widetilde{C}_{\hat{\m}\hat{\n}\hat{\r}}=i\frac{3}{2}\overline{\ve}\G_{\hat{\m}\hat{\n}}\p_{\hat{\r}}
\ea 
where the super covariant field strength is $\widetilde{F}_{\hat{\m}\hat{\n}\hat{\r}\hat{\s}}=\mathcal{\widetilde{F}}_{\hat{\m}\hat{\n}\hat{\r}\hat{\s}}-3i\pb_{\hat{\m}}\G_{\hat{\n}\hat{\r}}\p_{\hat{\s}}$.\\ 
As explained before, this method works as long as the signature of the space--time does not enter the computation; otherwise some more considerations are needed. Hence, performing a supersymmetry transformation on the Lagrangian, the  Chern--Simons term must cancel with the Pauli term, but  this involves the dualization of Lorentz indices on the Gamma--matrices, which depends on the number of spacelike dimensions.\\ When the number $s$ of spacelike dimensions is even, as in the standard $D=11$ supergravity, we have the usual definition:
\be\label{dualize}\e_{\a_1\dots\a_n\b_1\dots\b_m}\G^{\b_1\dots\b_m}=(-1)^{\frac{m(m-1)}{2}}(-i)m!\ \G_{\a_1\dots\a_n}\ee    
When $s$ is odd, as in the present case, equation \eq{dualize} must be modified with an extra ``$i$'' factor, that is:
\be\label{dualizex}\e_{\a_1\dots\a_n\b_1\dots\b_m}\G^{\b_1\dots\b_m}=(-1)^{\frac{m(m-1)}{2}}(-1)m!\ \G_{\a_1\dots\a_n}\ee 
This implies that we need an extra ''$i$'' also in the Chern--Simons term, in order to have cancellation.   
\ba
(det\,V)^{-1}\mathcal{L}&=&-\frac{1}{4}\mathcal{R}-\frac{i}{2}\pb_{\hat{\m}}\G^{\hat{\m}\hat{\n}\hat{\r}}\mathcal{D}_{\hat{\n}}\p_{\hat{\r}}+\frac{1}{48}\mathcal{\widetilde{F}}_{\hat{\m}\hat{\n}\hat{\r}\hat{\s}}\mathcal{\widetilde{F}}^{\hat{\m}\hat{\n}\hat{\r}\hat{\s}}-\frac{i}{192}\left(\pb_{\hat{\m}}\G^{\hat{\m}\hat{\n}\hat{\r}\hat{\s}\hat{\l}\hat{\pi}}\p_{\hat{\n}}+12\pb^{\hat{\r}}\G^{\hat{\s}\hat{\l}}\p^{\hat{\pi}}\right)\nn\\\ &&\left(\mathcal{\widetilde{F}}_{\hat{\r}\hat{\s}\hat{\l}\hat{\pi}}+\widetilde{F}_{\hat{\r}\hat{\s}\hat{\l}\hat{\pi}}\right)-\frac{2}{(12)^{4}}(det\,V)^{-1}\e^{\hat{\a}\hat{\b}\hat{\g}\hat{\d}\hat{\l}\hat{\m}\hat{\n}\hat{\pi}\hat{\s}\hat{\t}\hat{\o}}\mathcal{\widetilde{F}}_{\hat{\a}\hat{\b}\hat{\g}\hat{\d}}\mathcal{\widetilde{F}}_{\hat{\l}\hat{\m}\hat{\n}\hat{\pi}}\widetilde{C}_{\hat{\s}\hat{\t}\hat{\o}}\label{L11g}\ea
As we stressed before, we can see that the field $\mathcal{\widetilde{F}}_{\hat{\l}\hat{\m}\hat{\n}\hat{\pi}}$ is a ghost.
\subsection{The $D=10$ Type $IIA$ supergravities}
We will now consider the Type $IIA$ theories, starting from the already known $IIA(1,9)$ \cite{Campbell:zc}, to which we refer for notations and conventions; from the previous analysis we know that the $IIA(1,9)$, $IIA(5,5)$ and $IIA^*(9,1)$ have the same structure, that we can take from reference \cite{Campbell:zc}.

The supersymmetry transformation laws are (up to three fermion terms): 
\ba&&\d V^a_{\m}=-i\overline{\ve}\g^a\p_{\mu}\\
&&\d A_{\m}=\frac{i}{2}e^{-\frac{9}{8}\s}\pb_{\m}\g^{11}\ve-i\frac{3\sqrt{2}}{8}e^{-\frac{9}{8}\s}\cb\g_{\m}\ve\\
&&\d B_{\m\n}=e^{\frac{3}{4}\s}\pb_{[\m}\g_{\n]}\g^{11}\ve-\frac{1}{2\sqrt{2}}e^{\frac{3}{4}\s}\cb\g_{\m\n}\ve\\
&&\d C_{\m\n\r}=-\frac{3}{2}e^{-\frac{3}{8}\s}\pb_{[\m}\g_{\n\r]}\ve+\frac{1}{4\sqrt{2}}e^{-\frac{3}{8}\s}\cb\g^{11}\g_{\m\n\r}\ve+\nn\\ &&+6e^{\frac{3}{4}\s}A_{[\m}\pb_{\n}\g_{\r]}\g^{11}\ve-\frac{3}{\sqrt{2}}e^{\frac{3}{4}\s}A_{[\m}\cb\g_{\n\r]}\ve\\
&&\d\s=i\frac{2\sqrt{2}}{3}\cb\g^{11}\ve\\
&&\d\p_{\m}=\mathcal{D}_{\m}\ve-\frac{1}{32}e^{\frac{9}{8}\s}(\g_{\m}^{\phantom{D}\n\r}-14\d^{\n}_{\m}\g^{\r})\g^{11}G_{\n\r}\ve+\frac{i}{48}e^{-\frac{3}{4}\s}(\g_{\m}^{\phantom{D}\n\r\s}-9\d^{\n}_{\m}\g^{\r\s})\g^{11}H_{\n\r\s}\ve+\nn\\
&&+\frac{i}{128}e^{\frac{3}{8}\s}(\g_{\m}^{\phantom{D}\n\r\s\t}-\frac{20}{3}\d^{\n}_{\m}\g^{\r\s\t})F_{\n\r\s\t}\ve\\
&&\d\c=\frac{3\sqrt{2}}{8}\mathcal{D}_{\m}\s\g^{\m}\g^{11}\ve+\frac{3}{8\sqrt{2}}e^{\frac{9}{8}\s}\g^{\m\n}G_{\m\n}\ve+\frac{i}{12\sqrt{2}}e^{-\frac{3}{4}\s}\g^{\m\n\r}H_{\m\n\r}\ve+\nn\\
&&-\frac{i}{96\sqrt{2}}e^{\frac{3}{8}\s}\g^{\m\n\r\s}\g^{11}F_{\m\n\r\s}\ve\ea
The Lagrangian is given by (up to four fermions terms):
\ba&&(det\,V)^{-1}\mathcal{L}=-\frac{1}{4}\mathcal{R}-\frac{i}{2}\pb_{\m}\g^{\m\n\r}\mathcal{D}_{\n}\p_{\r}-\frac{1}{48}e^{\frac{3}{4}\s}\mathcal{F}_{\m\n\r\s}\mathcal{F}^{\m\n\r\s}+\frac{1}{12}e^{-\frac{3}{2}\s}\mathcal{H}_{\m\n\r}\mathcal{H}^{\m\n\r}+\nn\\&&-\frac{1}{4}e^{\frac{9}{4}\s}\mathcal{G}_{\m\n}\mathcal{G}^{\m\n}+\frac{i}{2}\cb\g^{\m}\mathcal{D}_{\m}\c-i\frac{3\sqrt{2}}{8}\cb\g^{11}\g^{\m}\g^{\n}\p_{\m}\partial_{\n}\s+\frac{9}{32}\partial_{\m}\s\partial^{\m}\s+\nn\\&&+\frac{3}{2(12)^3}(det\,V)^{-1}\e^{\a\b\g\d\l\m\n\r\s\t}\mathcal{F}_{\a\b\g\d}\mathcal{F}_{\l\m\n\r}B_{\s\t}+\nn\\
&&+\frac{1}{96}e^{\frac{3}{8}\s}(\pb_{\m}\g^{\m\n\l\r\s\t}\p_{\n}+12\pb^{\l}\g^{\r\s}\p^{\t}+\frac{1}{\sqrt{2}}\cb\g^{11}\g^{\m}\g^{\l\r\s\t}\p_{\m}+\frac{3}{4}\cb\g^{\l\r\s\t}\c)\mathcal{F}_{\l\r\s\t}+\nn\\
&&-\frac{1}{24}e^{-\frac{3}{4}\s}(\pb_{\m}\g^{11}\g^{\m\n\l\r\s}\p_{\n}-6\pb^{\l}\g^{11}\g^{\r}\p^{\s}-\sqrt{2}\cb\g^{\m}\g^{\l\r\s}\p_{\m})\mathcal{H}_{\l\r\s}+\nn\\
&&-\frac{i}{8}e^{\frac{9}{8}\s}(\pb_{\m}\g^{11}\g^{\m\n\l\r}\p_{\n}+2\pb^{\l}\g^{11}\p^{\r}+\frac{3}{\sqrt{2}}\cb\g^{\m}\g^{\l\r}\p_{\m}+\frac{5}{4}\cb\g^{11}\g^{\l\r}\c)\mathcal{G}_{\l\r}\label{L10nnn}\ea
where we have defined, according to reference \cite{Campbell:zc}, the super covariant field strength as
\ba\label{ef}&&F_{\m\n\r\s}=\mathcal{F}_{\m\n\r\s}-3e^{-\frac{3}{8}\s}\pb_{[\m}\g_{\n\r}\p_{\s]}+\frac{1}{\sqrt{2}}e^{-\frac{3}{8}\s}\cb\g^{11}\g_{[\m\n\r}\p_{\s]}\\
&&H_{\m\n\r}=\mathcal{H}_{\m\n\r}-\frac{3}{2}e^{\frac{3}{4}\s}\pb_{[\m}\g_{\n}\g^{11}\p_{\r]}+\frac{3}{2\sqrt{2}}e^{\frac{3}{4}\s}\cb\g_{[\m\n}\p_{\r]}\\
&&G_{\m\n}=\mathcal{G}_{\m\n}+\frac{i}{2}e^{-\frac{9}{8}\s}\pb_{\m}\g^{11}\p_{\n}-i\frac{3\sqrt{2}}{4}e^{-\frac{9}{8}\s}\cb\g_{[\m}\p_{\n]}\\&&\label{sig}\mathcal{D}_{\m}\s=\partial_{\m}\s+i\frac{2\sqrt{2}}{3}\pb_{\m}\g^{11}\c\ea 
and the field strength as
\be\label{sufi}\mathcal{F}_{\m\n\r\s}=4\partial_{[\m}C_{\m\n\r}+8A_{[\m}\mathcal{H}_{\n\r\s]};\quad\mathcal{H}_{\m\n\r}=3\partial_{[\m}B_{\n\r]};\quad\mathcal{G}_{\m\n}=2\partial_{[\m}A_{\n]}\ee
As discussed in the previous section, all the other theories have formally the same structure, but they contain some pure imaginary fields and we need to redefine them in terms of real ones case by case, according to the analysis of Table \eq{10curr2} and Table \eq{ghostfields} and the Appendix.\\For the theories $IIA^*(1,9)$, $IIA^*(5,5)$ and $IIA(9,1)$ we need to redefine\ba&&F=-i\widetilde{F};\quad\mathcal{F}=-i\widetilde{\mathcal{F}};\quad C=-i\widetilde{C}\\
&&G=-i\widetilde{G};\quad\mathcal{G}=-i\widetilde{\mathcal{G}};\quad A=-i\widetilde{A}\\
&&\c=i\widetilde{\c};\quad\cb=-i\widetilde{\cb}\ea  
where we have maintained the definition \eq{sufi} for the field strength. Thus, the supersymmetry transformation laws and the Lagrangian of theories $IIA^*(1,9)$, $IIA^*(5,5)$ and $IIA(9,1)$ are given by:
\ba&&\d V^a_{\m}=-i\overline{\ve}\g^a\p_{\mu}\\
&&\d\widetilde{A}_{\m}=-\frac{1}{2}e^{-\frac{9}{8}\s}\pb_{\m}\g^{11}\ve-i\frac{3\sqrt{2}}{8}e^{-\frac{9}{8}\s}\widetilde{\cb}\g_{\m}\ve\\
&&\d B_{\m\n}=e^{\frac{3}{4}\s}\pb_{[\m}\g_{\n]}\g^{11}\ve+\frac{i}{2\sqrt{2}}e^{\frac{3}{4}\s}\cb\g_{\m\n}\ve\\
&&\d\widetilde{C}_{\m\n\r}=-i\frac{3}{2}e^{-\frac{3}{8}\s}\pb_{[\m}\g_{\n\r]}\ve+\frac{1}{4\sqrt{2}}e^{-\frac{3}{8}\s}\widetilde{\cb}\g^{11}\g_{\m\n\r}\ve+\nn\\ &&+6e^{\frac{3}{4}\s}\widetilde{A}_{[\m}\pb_{\n}\g_{\r]}\g^{11}\ve+i\frac{3}{\sqrt{2}}e^{\frac{3}{4}\s}\widetilde{A}_{[\m}\cb\g_{\n\r]}\ve\\
&&\d\s=\frac{2\sqrt{2}}{3}\cb\g^{11}\ve\\
&&\d\p_{\m}=\mathcal{D}_{\m}\ve+\frac{i}{32}e^{\frac{9}{8}\s}(\g_{\m}^{\phantom{D}\n\r}-14\d^{\n}_{\m}\g^{\r})\g^{11}G_{\n\r}\ve+\frac{i}{48}e^{-\frac{3}{4}\s}(\g_{\m}^{\phantom{D}\n\r\s}-9\d^{\n}_{\m}\g^{\r\s})\g^{11}H_{\n\r\s}\ve+\nn\\
&&+\frac{1}{128}e^{\frac{3}{8}\s}(\g_{\m}^{\phantom{D}\n\r\s\t}-\frac{20}{3}\d^{\n}_{\m}\g^{\r\s\t})F_{\n\r\s\t}\ve\\
&&\d\c=-i\frac{3\sqrt{2}}{8}\mathcal{D}_{\m}\s\g^{\m}\g^{11}\ve-\frac{3}{8\sqrt{2}}e^{\frac{9}{8}\s}\g^{\m\n}G_{\m\n}\ve+\frac{1}{12\sqrt{2}}e^{-\frac{3}{4}\s}\g^{\m\n\r}H_{\m\n\r}\ve+\nn\\
&&+\frac{i}{96\sqrt{2}}e^{\frac{3}{8}\s}\g^{\m\n\r\s}\g^{11}F_{\m\n\r\s}\ve\ea
\ba&&(det\,V)^{-1}\mathcal{L}=-\frac{1}{4}\mathcal{R}-\frac{i}{2}\pb_{\m}\g^{\m\n\r}\mathcal{D}_{\n}\p_{\r}+\frac{1}{48}e^{\frac{3}{4}\s}\mathcal{\widetilde{F}}_{\m\n\r\s}\mathcal{\widetilde{F}}^{\m\n\r\s}+\frac{1}{12}e^{-\frac{3}{2}\s}\mathcal{H}_{\m\n\r}\mathcal{H}^{\m\n\r}+\nn\\&&+\frac{1}{4}e^{\frac{9}{4}\s}\mathcal{\widetilde{G}}_{\m\n}\mathcal{\widetilde{G}}^{\m\n}+\frac{i}{2}\widetilde{\cb}\g^{\m}\mathcal{D}_{\m}\widetilde{\c}-\frac{3\sqrt{2}}{8}\widetilde{\cb}\g^{11}\g^{\m}\g^{\n}\p_{\m}\partial_{\n}\s+\frac{9}{32}\partial_{\m}\s\partial^{\m}\s+\nn\\&&-\frac{3}{2(12)^3}(det\,V)^{-1}\e^{\a\b\g\d\l\m\n\r\s\t}\mathcal{\widetilde{F}}_{\a\b\g\d}\mathcal{\widetilde{F}}_{\l\m\n\r}B_{\s\t}+\nn\\
&&-\frac{i}{96}e^{\frac{3}{8}\s}(\pb_{\m}\g^{\m\n\l\r\s\t}\p_{\n}+12\pb^{\l}\g^{\r\s}\p^{\t}-\frac{i}{\sqrt{2}}\widetilde{\cb}\g^{11}\g^{\m}\g^{\l\r\s\t}\p_{\m}+\frac{3}{4}\widetilde{\cb}\g^{\l\r\s\t}\widetilde{\c})\mathcal{F}_{\l\r\s\t}+\nn\\
&&-\frac{1}{24}e^{-\frac{3}{4}\s}(\pb_{\m}\g^{11}\g^{\m\n\l\r\s}\p_{\n}-6\pb^{\l}\g^{11}\g^{\r}\p^{\s}+i\sqrt{2}\widetilde{\cb}\g^{\m}\g^{\l\r\s}\p_{\m})\mathcal{H}_{\l\r\s}+\nn\\
&&-\frac{1}{8}e^{\frac{9}{8}\s}(\pb_{\m}\g^{11}\g^{\m\n\l\r}\p_{\n}+2\pb^{\l}\g^{11}\p^{\r}-i\frac{3}{\sqrt{2}}\widetilde{\cb}\g^{\m}\g^{\l\r}\p_{\m}+\frac{5}{4}\widetilde{\cb}\g^{11}\g^{\l\r}\widetilde{\c})\mathcal{G}_{\l\r}\label{L10gng}\ea
Comparing the Lagrangian \eq{L10gng} with the Lagrangian \eq{L10nnn} we see that $\mathcal{\widetilde{F}}_{\m\n\r\s}$ and $\mathcal{\widetilde{G}}_{\m\n}$ are ghost, as previously discussed.\\
Furthermore the super covariant field strength are now defined as 
\ba&&\widetilde{F}_{\m\n\r\s}=\mathcal{\widetilde{F}}_{\m\n\r\s}-3ie^{-\frac{3}{8}\s}\pb_{[\m}\g_{\n\r}\p_{\s]}+\frac{1}{\sqrt{2}}e^{-\frac{3}{8}\s}\cb\g^{11}\g_{[\m\n\r}\p_{\s]}\\
&&H_{\m\n\r}=\mathcal{H}_{\m\n\r}-\frac{3}{2}e^{\frac{3}{4}\s}\pb_{[\m}\g_{\n}\g^{11}\p_{\r]}+\frac{3}{2\sqrt{2}}e^{\frac{3}{4}\s}\cb\g_{[\m\n}\p_{\r]}\\
&&\widetilde{G}_{\m\n}=\mathcal{\widetilde{G}}_{\m\n}-\frac{1}{2}e^{-\frac{9}{8}\s}\pb_{\m}\g^{11}\p_{\n}-i\frac{3\sqrt{2}}{4}e^{-\frac{9}{8}\s}\cb\g_{[\m}\p_{\n]}\\&&\mathcal{D}_{\m}\s=\partial_{\m}\s-\frac{2\sqrt{2}}{3}\pb_{\m}\g^{11}\widetilde{\c}\ea 

For the $IIA(0,10)$ and $IIA'(0,10)$ series, since we have an even number of spacelike dimensions, we must recall that the dualization rules are different, in order to write the correct Chern--simons term.\\ If the number $s$ of spacelike dimensions is odd, as in the the previous cases we have\be\label{dualize2}\e_{\a_1\dots\a_n\b_1\dots\b_m}\g^{\b_1\dots\b_m}=(-1)^{\frac{m(m-1)}{2}}(-i)m!\ \g^{11}\g_{\a_1\dots\a_n}\ee When $s$ is even, as in the present case, equation \eq{dualize2} must be modified with an extra ``$i$'' factor, that is:
\be\e_{\a_1\dots\a_n\b_1\dots\b_m}\g^{\b_1\dots\b_m}=(-1)^{\frac{m(m-1)}{2}}(-1)m!\ \g^{11}\g_{\a_1\dots\a_n}\ee 
thus we need an extra ''$i$'' also the Chern--Simons term, in order to have cancellation when performing a supersymmetry transformation on the Lagrangian.   

For the theories $IIA(0,10)$, $IIA(4,6)$, $IIA(8,2)$, $IIA'(10,0)$, $IIA'(6,4)$ and $IIA'(2,8)$, according to the discussion in the previous section we redefine\ba
&&H=i\widetilde{H};\quad\mathcal{H}=i\widetilde{\mathcal{H}};\quad B=i\widetilde{B}\\
&&G=-i\widetilde{G};\quad\mathcal{G}=-i\widetilde{\mathcal{G}};\quad A=-i\widetilde{A}\\
&&\c=i\widetilde{\c};\quad\cb=-i\widetilde{\cb}
\ea
where we have maintained the definition \eq{sufi} also for the ``tilded'' field strength.\\
The supersymmetry transformation laws and the Lagrangian are given by:
\ba&&\d V^a_{\m}=-i\overline{\ve}\g^a\p_{\mu}\\
&&\d\widetilde{A}_{\m}=-\frac{1}{2}e^{-\frac{9}{8}\s}\pb_{\m}\g^{11}\ve-i\frac{3\sqrt{2}}{8}e^{-\frac{9}{8}\s}\cb\g_{\m}\ve\\
&&\d\widetilde{B}_{\m\n}=-ie^{\frac{3}{4}\s}\pb_{[\m}\g_{\n]}\g^{11}\ve+\frac{1}{2\sqrt{2}}e^{\frac{3}{4}\s}\widetilde{\cb}\g_{\m\n}\ve\\
&&\d C_{\m\n\r}=-\frac{3}{2}e^{-\frac{3}{8}\s}\pb_{[\m}\g_{\n\r]}\ve-i\frac{1}{4\sqrt{2}}e^{-\frac{3}{8}\s}\widetilde{\cb}\g^{11}\g_{\m\n\r}\ve+\nn\\ &&-6ie^{\frac{3}{4}\s}\widetilde{A}_{[\m}\pb_{\n}\g_{\r]}\g^{11}\ve+\frac{3}{\sqrt{2}}e^{\frac{3}{4}\s}\widetilde{A}_{[\m}\widetilde{\cb}\g_{\n\r]}\ve\\
&&\d\s=\frac{2\sqrt{2}}{3}\widetilde{\cb}\g^{11}\ve\\
&&\d\p_{\m}=\mathcal{D}_{\m}\ve+\frac{i}{32}e^{\frac{9}{8}\s}(\g_{\m}^{\phantom{D}\n\r}-14\d^{\n}_{\m}\g^{\r})\g^{11}\widetilde{G}_{\n\r}\ve-\frac{1}{48}e^{-\frac{3}{4}\s}(\g_{\m}^{\phantom{D}\n\r\s}-9\d^{\n}_{\m}\g^{\r\s})\g^{11}\widetilde{H}_{\n\r\s}\ve+\nn\\
&&+\frac{i}{128}e^{\frac{3}{8}\s}(\g_{\m}^{\phantom{D}\n\r\s\t}-\frac{20}{3}\d^{\n}_{\m}\g^{\r\s\t})F_{\n\r\s\t}\ve\\
&&\d\widetilde{\c}=-i\frac{3\sqrt{2}}{8}\mathcal{D}_{\m}\s\g^{\m}\g^{11}\ve-\frac{3}{8\sqrt{2}}e^{\frac{9}{8}\s}\g^{\m\n}\widetilde{G}_{\m\n}\ve+\frac{i}{12\sqrt{2}}e^{-\frac{3}{4}\s}\g^{\m\n\r}\widetilde{H}_{\m\n\r}\ve+\nn\\
&&-\frac{1}{96\sqrt{2}}e^{\frac{3}{8}\s}\g^{\m\n\r\s}\g^{11}F_{\m\n\r\s}\ve\ea
\ba&&(det\,V)^{-1}\mathcal{L}=-\frac{1}{4}\mathcal{R}-\frac{i}{2}\pb_{\m}\g^{\m\n\r}\mathcal{D}_{\n}\p_{\r}-\frac{1}{48}e^{\frac{3}{4}\s}\mathcal{F}_{\m\n\r\s}\mathcal{F}^{\m\n\r\s}-\frac{1}{12}e^{-\frac{3}{2}\s}\mathcal{\widetilde{H}}_{\m\n\r}\mathcal{\widetilde{H}}^{\m\n\r}+\nn\\&&+\frac{1}{4}e^{\frac{9}{4}\s}\mathcal{\widetilde{G}}_{\m\n}\mathcal{\widetilde{G}}^{\m\n}+\frac{i}{2}\widetilde{\cb}\g^{\m}\mathcal{D}_{\m}\widetilde{\c}-\frac{3\sqrt{2}}{8}\widetilde{\cb}\g^{11}\g^{\m}\g^{\n}\p_{\m}\partial_{\n}\s+\frac{9}{32}\partial_{\m}\s\partial^{\m}\s+\nn\\&&-\frac{3}{2(12)^3}(det\,V)^{-1}\e^{\a\b\g\d\l\m\n\r\s\t}\mathcal{F}_{\a\b\g\d}\mathcal{F}_{\l\m\n\r}\widetilde{B}_{\s\t}+\nn\\
&&+\frac{1}{96}e^{\frac{3}{8}\s}(\pb_{\m}\g^{\m\n\l\r\s\t}\p_{\n}+12\pb^{\l}\g^{\r\s}\p^{\t}-i\frac{1}{\sqrt{2}}\widetilde{\cb}\g^{11}\g^{\m}\g^{\l\r\s\t}\p_{\m}+\frac{3}{4}\widetilde{\cb}\g^{\l\r\s\t}\widetilde{\c})\mathcal{F}_{\l\r\s\t}+\nn\\
&&-\frac{i}{24}e^{-\frac{3}{4}\s}(\pb_{\m}\g^{11}\g^{\m\n\l\r\s}\p_{\n}-6\pb^{\l}\g^{11}\g^{\r}\p^{\s}+i\sqrt{2}\widetilde{\cb}\g^{\m}\g^{\l\r\s}\p_{\m})\mathcal{\widetilde{H}}_{\l\r\s}+\nn\\
&&-\frac{1}{8}e^{\frac{9}{8}\s}(\pb_{\m}\g^{11}\g^{\m\n\l\r}\p_{\n}+2\pb^{\l}\g^{11}\p^{\r}-i\frac{3}{\sqrt{2}}\widetilde{\cb}\g^{\m}\g^{\l\r}\p_{\m}+\frac{5}{4}\widetilde{\cb}\g^{11}\g^{\l\r}\widetilde{\c})\mathcal{\widetilde{G}}_{\l\r}\label{L10ngg}\ea
Comparing the Lagrangian \eq{L10ngg} with the Lagrangian \eq{L10nnn} we see that $\mathcal{\widetilde{H}}_{\m\n\r}$ and $\mathcal{\widetilde{G}}_{\m\n}$ are ghost, as previously discussed.\\
Here we have defined the super covariant field strength as
\ba&&F_{\m\n\r\s}=\mathcal{F}_{\m\n\r\s}-3e^{-\frac{3}{8}\s}\pb_{[\m}\g_{\n\r}\p_{\s]}-i\frac{1}{\sqrt{2}}e^{-\frac{3}{8}\s}\widetilde{\cb}\g^{11}\g_{[\m\n\r}\p_{\s]}\\
&&\widetilde{H}_{\m\n\r}=\mathcal{\widetilde{H}}_{\m\n\r}+i\frac{3}{2}e^{\frac{3}{4}\s}\widetilde{\pb}_{[\m}\g_{\n}\g^{11}\p_{\r]}-\frac{3}{2\sqrt{2}}e^{\frac{3}{4}\s}\cb\g_{[\m\n}\p_{\r]}\\
&&\widetilde{G}_{\m\n}=\mathcal{\widetilde{G}}_{\m\n}-\frac{1}{2}e^{-\frac{9}{8}\s}\pb_{\m}\g^{11}\p_{\n}-i\frac{3\sqrt{2}}{4}e^{-\frac{9}{8}\s}\widetilde{\cb}\g_{[\m}\p_{\n]}\\&&\mathcal{D}_{\m}\s=\partial_{\m}\s-\frac{2\sqrt{2}}{3}\pb_{\m}\g^{11}\widetilde{\c}\ea 
Finally, we derive the structure of $IIA(10,0)$, $IIA(6,4)$, $IIA(2,8)$, $IIA'(0,10)$, $IIA'(4,6)$ and $IIA'(8,2)$, for which we define   
\ba&&F=-i\widetilde{F};\quad\mathcal{F}=-i\widetilde{\mathcal{F}};\quad C=-i\widetilde{C}\\
&&H=-i\widetilde{H};\quad\mathcal{H}=-i\widetilde{\mathcal{H}};\quad B=-i\widetilde{B}\ea
maintaining the definition \eq{sufi} for the ``tilded'' field strength.\\
The supersymmetry transformation laws and the Lagrangian are given by:
\ba&&\d V^a_{\m}=-i\overline{\ve}\g^a\p_{\mu}\\
&&\d A_{\m}=\frac{i}{2}e^{-\frac{9}{8}\s}\pb_{\m}\g^{11}\ve-i\frac{3\sqrt{2}}{8}e^{-\frac{9}{8}\s}\cb\g_{\m}\ve\\
&&\d\widetilde{B}_{\m\n}=ie^{\frac{3}{4}\s}\pb_{[\m}\g_{\n]}\g^{11}\ve-\frac{i}{2\sqrt{2}}e^{\frac{3}{4}\s}\cb\g_{\m\n}\ve\\
&&\d\widetilde{C}_{\m\n\r}=-i\frac{3}{2}e^{-\frac{3}{8}\s}\pb_{[\m}\g_{\n\r]}\ve+i\frac{1}{4\sqrt{2}}e^{-\frac{3}{8}\s}\cb\g^{11}\g_{\m\n\r}\ve+\nn\\ &&+6ie^{\frac{3}{4}\s}A_{[\m}\pb_{\n}\g_{\r]}\g^{11}\ve-i\frac{3}{\sqrt{2}}e^{\frac{3}{4}\s}A_{[\m}\cb\g_{\n\r]}\ve\\
&&\d\s=i\frac{2\sqrt{2}}{3}\cb\g^{11}\ve\\
&&\d\p_{\m}=\mathcal{D}_{\m}\ve-\frac{1}{32}e^{\frac{9}{8}\s}(\g_{\m}^{\phantom{D}\n\r}-14\d^{\n}_{\m}\g^{\r})\g^{11}G_{\n\r}\ve+\frac{1}{48}e^{-\frac{3}{4}\s}(\g_{\m}^{\phantom{D}\n\r\s}-9\d^{\n}_{\m}\g^{\r\s})\g^{11}\widetilde{H}_{\n\r\s}\ve+\nn\\
&&+\frac{1}{128}e^{\frac{3}{8}\s}(\g_{\m}^{\phantom{D}\n\r\s\t}-\frac{20}{3}\d^{\n}_{\m}\g^{\r\s\t})\widetilde{F}_{\n\r\s\t}\ve\\
&&\d\c=\frac{3\sqrt{2}}{8}\mathcal{D}_{\m}\s\g^{\m}\g^{11}\ve+\frac{3}{8\sqrt{2}}e^{\frac{9}{8}\s}\g^{\m\n}G_{\m\n}\ve+\frac{1}{12\sqrt{2}}e^{-\frac{3}{4}\s}\g^{\m\n\r}\widetilde{H}_{\m\n\r}\ve+\nn\\
&&-\frac{1}{96\sqrt{2}}e^{\frac{3}{8}\s}\g^{\m\n\r\s}\g^{11}\widetilde{F}_{\m\n\r\s}\ve\ea
\ba&&(det\,V)^{-1}\mathcal{L}=-\frac{1}{4}\mathcal{R}-\frac{i}{2}\pb_{\m}\g^{\m\n\r}\mathcal{D}_{\n}\p_{\r}+\frac{1}{48}e^{\frac{3}{4}\s}\mathcal{\widetilde{F}}_{\m\n\r\s}\mathcal{\widetilde{F}}^{\m\n\r\s}-\frac{1}{12}e^{-\frac{3}{2}\s}\mathcal{\widetilde{H}}_{\m\n\r}\mathcal{\widetilde{H}}^{\m\n\r}+\nn\\&&-\frac{1}{4}e^{\frac{9}{4}\s}\mathcal{G}_{\m\n}\mathcal{G}^{\m\n}+\frac{i}{2}\cb\g^{\m}\mathcal{D}_{\m}\c-i\frac{3\sqrt{2}}{8}\cb\g^{11}\g^{\m}\g^{\n}\p_{\m}\partial_{\n}\s+\frac{9}{32}\partial_{\m}\s\partial^{\m}\s+\nn\\&&-\frac{3}{2(12)^3}(det\,V)^{-1}\e^{\a\b\g\d\l\m\n\r\s\t}\mathcal{\widetilde{F}}_{\a\b\g\d}\mathcal{\widetilde{F}}_{\l\m\n\r}\widetilde{B}_{\s\t}+\nn\\
&&-\frac{i}{96}e^{\frac{3}{8}\s}(\pb_{\m}\g^{\m\n\l\r\s\t}\p_{\n}+12\pb^{\l}\g^{\r\s}\p^{\t}+\frac{1}{\sqrt{2}}\cb\g^{11}\g^{\m}\g^{\l\r\s\t}\p_{\m}+\frac{3}{4}\cb\g^{\l\r\s\t}\c)\mathcal{F}_{\l\r\s\t}+\nn\\
&&+\frac{i}{24}e^{-\frac{3}{4}\s}(\pb_{\m}\g^{11}\g^{\m\n\l\r\s}\p_{\n}-6\pb^{\l}\g^{11}\g^{\r}\p^{\s}-\sqrt{2}\cb\g^{\m}\g^{\l\r\s}\p_{\m})\mathcal{H}_{\l\r\s}+\nn\\
&&-\frac{i}{8}e^{\frac{9}{8}\s}(\pb_{\m}\g^{11}\g^{\m\n\l\r}\p_{\n}+2\pb^{\l}\g^{11}\p^{\r}+\frac{3}{\sqrt{2}}\cb\g^{\m}\g^{\l\r}\p_{\m}+\frac{5}{4}\cb\g^{11}\g^{\l\r}\c)\mathcal{G}_{\l\r}\label{L10ggn}\ea
Comparing the Lagrangian \eq{L10ggn} with the Lagrangian \eq{L10nnn} we see that $\mathcal{\widetilde{F}}_{\m\n\r\s}$ and $\mathcal{\widetilde{H}}_{\m\n\r}$ are ghost, as previously discussed.\\
Here we have defined the super covariant field strength as
\ba&&\widetilde{F}_{\m\n\r\s}=\mathcal{\widetilde{F}}_{\m\n\r\s}-3ie^{-\frac{3}{8}\s}\pb_{[\m}\g_{\n\r}\p_{\s]}+\frac{i}{\sqrt{2}}e^{-\frac{3}{8}\s}\cb\g^{11}\g_{[\m\n\r}\p_{\s]}\\
&&\widetilde{H}_{\m\n\r}=\mathcal{\widetilde{H}}_{\m\n\r}-i\frac{3}{2}e^{\frac{3}{4}\s}\pb_{[\m}\g_{\n}\g^{11}\p_{\r]}+i\frac{3}{2\sqrt{2}}e^{\frac{3}{4}\s}\cb\g_{[\m\n}\p_{\r]}\\
&&G_{\m\n}=\mathcal{G}_{\m\n}+\frac{i}{2}e^{-\frac{9}{8}\s}\pb_{\m}\g^{11}\p_{\n}-i\frac{3\sqrt{2}}{4}e^{-\frac{9}{8}\s}\cb\g_{[\m}\p_{\n]}\\&&\mathcal{D}_{\m}\s=\partial_{\m}\s+i\frac{2\sqrt{2}}{3}\pb_{\m}\g^{11}\c\ea

\section{Conclusions}
In this paper we have constructed with the geometric approach the supersymmetry transformation laws and Lagrangian for all the variant $D=11$ and $D=10$ Type $IIA$ supergravities and  shown that they admit a supersymmetric Poincar\'e vacuum. With our constructive approach we realized that for each allowed signature there exists two distinct $D=10$ Type $IIA$ supergravities. When variant supergravities were first introduced by Hull \cite{Hull:1998ym}, two versions of $D=10$ Type $IIA$ theories occurred only for signatures $(t,s)=(1,9)$, $(t,s)=(9,1)$, $(t,s)=(5,5)$. The theories classified by Hull  are in fact the only $D=10$ Type $IIA$ theories that can be obtained from $D=11$ supergravity compactifying on a spacelike or on a timelike circle. The set of new theories constructed in this paper can certainly not be obtained trivially compactifying from eleven dimensions on just one circle, and seem to be distinct from the other ones. It would be interesting to understand how they arise in the context of $T$--duality transformations.\\    
The procedure used in this context to construct variant forms of an already known supergravity, can be suitably extended to other cases.

\section*{Acknowledgments}
The author is grateful to Sergio Ferrara, Riccardo D'Auria and Laura Andrianopoli for useful suggestions and discussions. Work
supported in by the European Community's Human Potential
Program under contract HPRN-CT-2000-00131 Quantum Space-Time, in
which the author is associated to Torino University
\section*{Appendix}
\setcounter{equation}{0}
\addtocounter{section}{1}

In this appendix we discuss the consistency of the Majorana condition on spinors for various signatures and for the two definitions of the Dirac conjugate spinor $\pb=\p^{\dagger}G^{-1}$ $G=\{G_I,G_{II}\}$.\\
In $D=11$ the only possible choice for the charge conjugation matrix is the antisymmetric one $C^T=-C$, so that:
\be\label{trans}\g_a^T=-C\g_aC^{-1}\ee
On the other hand, the hermitian conjugate depends on the choice of $G$, since
\be\label{dagger3}\G^{a\dagger}=\eta\, G^{-1}\G^aG \ee
where the phase $\eta$ takes two different values according to the two definitions of $G$, namely 
\be\label{dagger4}\eta_I=(-1)^{t-1};\,\,\,\,\,\eta_{II}=(-1)^s\ee
We can observe that in odd dimensions, the phases \eq{dagger4} coincide, so that we have only one way to define the hermitian conjugate of a $\g$ matrix. On the other hand, in odd dimensions, $G_I$ and $G_{II}$ differ only by a phase factor \cite{Regge:1983uu}, depending on the signature of the space--time. That means that if this factor is real, the reality of a current evaluated with definitions $I$ and $II$ is the same in both cases; if the scalar factor is pure imaginary each current has opposite reality in the two cases, but this can be formally absorbed in the definition of $\d$ in equation \eq{coniugo}. Thus, in odd dimension, is totally equivalent to use definition $I$ or $II$ for the Dirac conjugate spinor; since we have chosen for our calculations $G=G_I$ we will consider this latter case and we impose the following Majorana condition on spinors:
\be\label{majo11}\p^{\dagger}G_I^{-1}=\p^TC\ee
this must be consistent with the supersymmetry transformation rules of such spinors
\be\label{varmajo11}\d\p^{\dagger}G_I^{-1}=\d\p^TC\ee
At this purpose, we can evaluate the transposition and the hermitian
conjugation of the product of $n$ Gamma--matrices in $D=11$ using
equations \eq{trans}, \eq{dagger3}, \eq{dagger4}:
\ba
&&(\g_{a_1}\dots\g_{a_n})^T=(-1)^nC_{(-)}^{-1}\g_{a_n}\dots\g_{a_1}C_{(-)}\nn\\
&&\label{rulez}(\g_{a_1}\dots\g_{a_n})^{\dagger}=(-1)^{n(t-1)} G_{I}^{-1}\g_{a_n}\dots\g_{a_1}G_{I}
\ea
If we want to satisfy the condition \eq{varmajo11}, the coefficients of the supersymmetry transformation rule of $\p$ \eq{due11ng},\eq{due11g} must obey a simple rule descending from equations \eq{rulez}: when
$t$ is even the product  must appear in the transformation rule
with a real coefficient; when $t$ is
odd, there must be a real coefficient if $n$ is even and an imaginary one if $n$ is odd. This is exactly what we found in Section 4.

These considerations become more interesting in the $D=10$ case. When real scalar fields are present they must appear, out of the vacuum configuration, as in equation \eq{sig}, that is\be\mathcal{D}_{\m}\s=\partial_{\m}\s+i\frac{2\sqrt{2}}{3}\pb_{\m}\g^{11}\c\ee In order to have the scalar field $\s$ real, the coefficient of $\pb_{\m}\g^{11}\c$  must be chosen according to the reality properties of the current that are summarized in the last line of Table \ref{10curr}\\ 
According to our previous discussions, we could be tempted to conclude that in the $IIA^*(1,9)$ and $IIA(0,10)$ series, the field $\s$ becomes pure imaginary and then it is a ghost. On the other hand we have to take into account the possibility that the spin $\frac{1}{2}$ field $\c$ should be redefined as $\c\longrightarrow i\c$. The right choice can be made considering the structure of the supersymmetry transformation laws of $\p$ and $\c$ and requiring that also $\d\p$ and $\d\c$ satisfy a Majorana condition (in the present case it very easy to see that the dilaton $\s$ could never become pure imaginary, because of its coupling $e^{\s}$).\\ In $D=10$ we have chosen the same charge conjugation matrix as in $D=11$, so we evaluate the transposition of the product of $n$ matrices $\g$ as in $D=11$; on the other hand, for hermitian conjugation, we must consider the two distinct definitions $I$, $II$ for the Dirac conjugate spinor. Using equations \eq{dagger3}, \eq{dagger4}, we can evaluate the hermitian conjugate of the product of $n$ Gamma--matrices in the two cases and compare them with the transposition: 
\ba
&&(\g_{a_1}\dots\g_{a_n})^T=(-1)^nC_{(-)}^{-1}\g_{a_n}\dots\g_{a_1}C_{(-)}\nn\\
&&I:\quad(\g_{a_1}\dots\g_{a_n})^{\dagger}=(-1)^{n(t-1)} G_{I}^{-1}\g_{a_n}\dots\g_{a_1}G_{I}\nn\\
&&II:\quad(\g_{a_1}\dots\g_{a_n})^{\dagger}=(-1)^{ns} G_{II}^{-1}\g_{a_n}\dots\g_{a_1}G_{II}\label{rulez10}
\ea
Also in this case it is easy to find a simple rule for the reality of the coefficients in front of  a term with $n$ Gamma--matrices in the supersymmetry transformation laws of the fermions. Using $G=G_I$, if $t$ is even the coefficient is always real; if $t$ is odd, it is real if $n$ is even, pure imaginary if $n$ is odd. Using $G=G_{II}$, if $s$ is odd the coefficient is always real; if $s$ is even, it is real if $n$ is even, pure imaginary if $n$ is odd.\\ 
Taking into account the various definition of the matrix $\g^{11}$ it is possible to identify $``a\, priori$''  the reality of the coefficients appearing in the supersymmetry transformation laws of the fermions for the various cases, in order to have Majorana spinors.
We see that we have Majorana spinors if in the $IIA^*(1,9)$ and $IIA(0,10)$ series we define $\c\longrightarrow i\c$, so that the dilaton never becomes a ghost.

\end{document}